\documentclass[prl,aps,10pt,superscriptaddress,twocolumn,floatfix,nofootinbib]{revtex4-1}
\usepackage{graphicx,color}
\usepackage[dvipsnames]{xcolor}
\usepackage{amsmath,amssymb,bm}
\usepackage[plainpages=false,pdfpagelabels,colorlinks=true,linkcolor=red,urlcolor=blue,citecolor=blue,
pdftitle={Effective Spin Foam Models for 4D Quantum Gravity},
pdfauthor={Seth K. Asante, Bianca Dittrich, Hal M. Haggard},
pdfdisplaydoctitle=true,pdfduplex=DuplexFlipLongEdge,bookmarks=false]{hyperref}

\usepackage{pgfplots}
\usepackage[normalem]{ulem}
\usepackage{subcaption}

\def\be#1\ee{\begin{align}#1\end{align}}

\def\ba{\begin{eqnarray}}
\def\ea{\end{eqnarray}}
\def\nn{\nonumber}
\def\q{\quad}

\begin{document}

\title{Effective Spin Foam Models for Four-Dimensional Quantum Gravity}

\author{Seth K. Asante}
\affiliation{Perimeter Institute, 31 Caroline Street North, Waterloo, ON, N2L 2Y5, CAN}
\affiliation{Department of Physics and  Astronomy, University of Waterloo, 200 University Avenue West,Waterloo, Ontario, Canada, N2L 3G1}
\author{Bianca Dittrich}
\affiliation{Perimeter Institute, 31 Caroline Street North, Waterloo, ON, N2L 2Y5, CAN}
\author{Hal M. Haggard}
\affiliation{Physics Program, Bard College, 30 Campus Road, Annandale-On-Hudson, NY 12504, USA}
\affiliation{Perimeter Institute, 31 Caroline Street North, Waterloo, ON, N2L 2Y5, CAN}

\begin{abstract}
 A number of approaches to four-dimensional quantum gravity, such as loop quantum gravity and holography,  situate areas as their fundamental variables. However, this choice of kinematics can easily lead to gravitational dynamics peaked on flat spacetimes. We show that this is due to how regions are glued in the gravitational path integral via a discrete spin foam model. We introduce a family of `effective' spin foam models that incorporate a quantum area spectrum, impose gluing constraints as strongly as possible, and leverage the discrete general relativity action to specify amplitudes. These effective spin foam models avoid flatness in a restricted regime of the parameter space.  
\end{abstract}

\maketitle

\emph{Symplectic, metrical quantization}.  Envisioning the geometry of spacetime as dynamically evolving founded the revolutionary insights of general relativity (GR) that have  resulted in direct measurements of gravitational time dilation, bending of starlight, and gravitational waves. However, this revolution remains incomplete. We still do not know how to fully characterize an evolving quantum spacetime geometry. 

The quantization of spacetime geometry is an interplay between its symplectic and metrical aspects. The former determines the allowed phase space and its associated quantum theory, while the latter encodes spacetime dynamics. In three dimensions (3D) alignment between these two facets of geometry allows one to 
construct a discrete, simplicial path integral for  quantum gravity, the Ponzano-Regge model \cite{PR}.  Spacetime is decomposed into a large collection of tetrahedra  that are glued along a subset of edges with matched lengths. Metrical and symplectic aspects of this geometry nicely align:  lengths encode the intrinsic metric and  tedrahedral dihedral angles  encode the extrinsic geometry and these variables are canonically conjugated to each other \cite{WaelbroeckZapata,BonzomDittrich}. In the Euclidean signature case the angles are compact,  which leads to discrete spectra for the lengths.

In 4D the situation is more subtle, and there is some  tension between the symplectic and the metrical.  In a space-time split, the metric has two natural discretizations:  the lengths of edges, 
and  the extrinsic curvature angles defined around 2D faces.

These variables are not canonically conjugate. This forces a choice: either the lengths or the extrinsic curvature angles must be completed to a set of canonically conjugated coordinates.

If the lengths are chosen, then the conjugate variables are contractions of the curvature angles with certain area-length derivatives  \cite{DittrichHoehn1}. These variables have, so far, resisted rigorous quantization.
 Meanwhile for the curvature angles, the conjugate variables are  the 2D face areas, whose quantization give a discrete area spectrum; this is because they are conjugate to compact angles. These variables arise  naturally in connection formulations of GR, like Loop Quantum Gravity (LQG) \cite{AshtekarVariables}. A key result of LQG is the rigorous quantization of area and volume observables, which indeed have discrete spectra  \cite{DiscreteGeom1,DiscreteGeom2,BianchiHaggard1,Wieland1}.

In spin foam models \cite{Perez,BC,NSFM}, which are discrete geometry path integrals derived from LQG, area variables are fundamental. Area variables play a central role in holography \cite{tHooft:1993dmi,SmolinHol}, in particular, for the reconstruction of geometry from entanglement \cite{RyuTakayanagi, BianchiMyers}, and discrete area spectra are key in many approaches to black hole entropy counting \cite{BHCounting,ABCK,Bek1,BarberoPerez}.  There is, however, an interplay between the choice of area variables
and the dynamics of GR: in this letter we show that area variables  must be constrained to avoid a suppression of curvature and that the discreteness of their spectra hinders sharp imposition of these constraints.

In fact, it has been argued that, in the semiclassical limit, flat configurations dominate the spin foam path integral \cite{FlatProblem,Bonzom1,HellK, Han1,Oliveira1, DGS, ILQGS}. We reveal the mechanism behind this unfortunate dominance, and identify a more favorable regime in which the path integral can peak on curved configurations. This `flatness problem' has been a key open problem for spin foams \cite{ILQGS}.  We show that it can be traced back to fundamental, discrete, area variables.

We tackle directly the question of whether a discrete, locally independent, area spectrum is consistent with the dynamics of GR. To this end we propose a family of `effective' models that (a) incorporate a discrete area spectrum, (b) impose the constraints between the areas as strongly as allowed by  the LQG Hilbert space structure, and (c) use---more directly than current spin foam models---a discretized GR action for the amplitudes.

These effective models allow us to show that the flatness problem can be overcome, but to do so also imposes certain restrictions involving the discretization scale,  curvature per triangle, and the Barbero-Immirzi parameter, which controls the area spectral spacing. Future work will show whether these restrictions are sufficient to ensure general relativistic dynamics in the continuum limit.

\medskip

\emph{Discrete, locally-independent areas}. We study a path integral for 4D quantum gravity regulated by a triangulation of spacetime. 
We work with quantum amplitudes for Euclidean signature, leaving the Lorentzian case to future work. Our key assumption is that the areas have a discrete, prescribed spectrum. We take the area eigenvalues to be independent, that is, their values will not depend on the state away from the measured triangle.
The particular area spectrum we work with is 
\be\label{ASpec}
a(j) = \gamma \ell_{P}^2 \sqrt{j(j+1)} \sim  \gamma \ell_{P}^2 ( j +1/2),
\ee
where $j$ is a half-integer spin label, $\ell_{P} = \sqrt{ 8\pi \hbar G/c^3}$ is the Planck length,  $\gamma$ is the dimensionless Barbero-Immirzi parameter, and $\sim$ indicates the large-$j$ asymptotic limit. We focus on the equispaced asymptotic spectrum.  This form for the area spectrum was established in LQG \cite{DiscreteGeom1,DiscreteGeom2,BianchiHaggard1,Wieland1}, but discrete areas have also been  discussed in the context of black hole spectroscopy \cite{BHCounting}. In LQG, triangle normals are represented by angular momentum vector operators rescaled by $\gamma$  \cite{Barbieri:1997ks,BianchiDonaSpeziale}. The area is thus given by the  square root of the SU(2) Casimir, multiplied by $\gamma$, which determines the area eigenvalue spacing.

Before taking up the path integral, we review the use of area variables in simplicial discretizations of GR. These discretizations were introduced
by Regge \cite{Regge} and used length variables. 
A wide array of reformulations have been considered \cite{AreaRegge, ReggeOther, BarrettFirstOrder, BahrDittrich, AreaAngle, ADH}, and we use descriptive adjectives to capture the variables used in each form. The change from length to area variables turns out to be more subtle than one might expect. A treatment in the more transparent context of Regge calculus will illuminate the issues before discussing the path integral. 

\medskip

\emph{Actions for discretized GR}. In Length Regge Calculus (LRC) one replaces the metric by lengths $l_e$ assigned to the edges $e$ of a triangulation. The $l_e$ determine the triangle areas $A_t(l)$  and the 4D (internal) dihedral angles $\theta^\sigma_t(l)$ in each 4-simplex $\sigma$. The action  
\ba\label{LRCA}
S_{\rm LRC} =    \sum_{t } n_t\pi A_t      - \sum_\sigma \sum_{t \supset \sigma} A_t \theta_t^\sigma
\equiv \sum_{t } S^l_t +
\sum_{\sigma} S^l_\sigma, \quad
\ea
is a discretization of the Einstein-Hilbert action and the corresponding equations of motion approximate Einstein's equations  \cite{Brewin:2000zh}. The factor $n_t \in \{1,2\}$ allows for triangulations with boundary and is 1 for triangles on the boundary and 2 for triangles in the bulk.

The 4-simplices, which are the basic building blocks of the triangulation, each have 10 edges and 10 areas. One can thus (locally) invert the 10 functions $A_t(l)$ that give a simplex's areas in terms of its lengths \cite{inversion}. We will denote the resulting functions $L^\sigma_e(a)$, where $a$ collectively signifies the 10 areas associated to $\sigma$.    This allows us to define the Area Regge Calculus (ARC) action \cite{AreaRegge,ReggeOther,ADH}, whose value on configurations with $a_t=A_t(l)$ agrees with the LRC action
\ba\label{ARegge}
S_{\rm ARC}&=& \sum_{t } \!S^a_t(a) +
\sum_{\sigma} \!S^a_\sigma(a) ,
\ea
where $S^a_t(a)=n_t\pi a_t$ and $S^a_\sigma(a)=S^l_\sigma(L^\sigma(a))$. Strikingly, freely varying the bulk areas one finds that the deficit angles $\epsilon_t=2\pi- \sum_{\sigma \supset t}\theta_t^\sigma $, which measure curvature, have to vanish \cite{ADH}. That is, the ARC equations of motion impose flatness.

Extended triangulations are built up by gluing pairs of 4-simplices $(\sigma,\sigma')$ through a shared tetrahedron $\tau$. Gluing identifies six pairs of length variables but only four pairs of area variables, which explains why the equations of motion for LRC and ARC differ: after gluing there are generically more triangles than edges in a 4D triangulation and thus more area than length variables. Restricting variation of the areas to a constraint surface coming from a consistent length assignment $a_t=A_t(l)$, one recovers the LRC equations of motion.

From this counting we see that working with area variables we miss two matching conditions per bulk tetrahedron. The geometry of a tetrahedron, however, can be uniquely specified by its four areas {\it and} two 3D dihedral angles at non-opposite edges (intriguingly, opposite dihedral angles do not suffice).  Introducing the 3D dihedral angles $\Phi_e^{\tau,\sigma}(a)=\Phi_e^\tau(L^\sigma(a))$, we have thus, two constraints per bulk tetrahedron
\ba\label{match1}
 \Phi_{e_i}^{\tau,\sigma}(a) - \Phi_{e_i}^{\tau,\sigma'}(a)  \,\stackrel{!}{=}\,0\, \q i=1,2,
\ea
where $(e_1,e_2)$ is a choice of a pair of non-opposite edges in $\tau$. Together with the matched areas, these constraints ensure that the lengths of a shared tetrahedron, as defined by the areas associated to $\sigma$ and $\sigma'$, match.

The constraints (\ref{match1}) involve pairs of neighbouring simplices. Introducing two 3D dihedral angles $\phi_{e_i}^\tau$ per tetrahedron as additional variables \cite{AreaAngle}, we can formulate alternative constraints, localized on a given 4-simplex $\sigma$:
\ba\label{match2}
\phi^\tau_{e_i} - \Phi_{e_i}^{\tau,\sigma}(a)  \,\stackrel{!}{=}\,0\,\, \q i=1,2.
\ea
These constraints fix all variables $\phi_{e_i}^\tau$ as functions of the areas {\it and} impose the constraints (\ref{match1}). We favor these localized constraints because they preserve the additive factorization of the action (\ref{ARegge}) and lead to the simplifying product factorization of the path integral below, Eq. (\ref{NewPI1}).

In the quantum theory, areas are encoded using SU(2) representation labels $j_t$, cf. Eq. (\ref{ASpec}),  which result from identifying triangle normals with angular momentum operators. The 3D dihedral angles are given by the inner product of these normals, and can be encoded in the recoupling of two angular momenta. The angles at a pair of non-opposite edges $(e_1,e_2)$ in a tetrahedron $\tau$ require different recoupling schemes and are therefore non-commutative \cite{DittrichRyan,noncommutative} (for a simplified proof see \cite{Supp}):
\ba\label{PB1}
\hbar \{\phi^\tau_{e_1},\phi^
{\tau}_{e_2}\}= \ell^2_P \gamma \frac{\sin \alpha^{t,\tau}_v}{a_t}= \frac{\sin \alpha^{t,\tau}_v}{(j_t+\tfrac{1}{2})}  \, ,
\ea
where $\alpha^{t,\tau}_v$ is the angle between $(e_1,e_2)$. Thus, the constraints (\ref{match2})  are also non-commutative,  more precisely, second class. For  these second class constraints the uncertainty relations prevent a sharp imposition of the constraints in the quantum theory. Armed with these understandings, we take up the path integral.

\medskip

\emph{Path integral}. 
To  incorporate a discrete area spectrum (\ref{ASpec}), we employ constrained ARC, and sum over spin labels $j_t$:
\ba\label{Z1}
{\cal Z}&=&\sum_{\{j_t\}} \mu(j) \prod_{t } {\cal A}_t(j) \prod_\sigma {\cal A}_\sigma(j) \prod_{\tau \in \text{blk}} G^{\sigma,\sigma'}_\tau (j) \, .
\ea
The triangle   ${\cal A}_{t } =\exp( {\rm i} \gamma  n_t   \pi (j_t+\tfrac{1}{2}))$  and simplex amplitudes ${\cal A}_\sigma=\exp\left( -{\rm i} \gamma\sum_{t\in \sigma} (j_t+\tfrac{1}{2}) \theta_t^\sigma(j)\right)$ result from the exponentiated ARC action (\ref{ARegge}). The precise form of the measure factor $\mu(j)$ will not be important for the discussion here, but see \cite{PImeasure}.
The factors $G^{\sigma,\sigma'}_\tau$ implement the constraints (\ref{match1}), and are crucial for imposing the dynamics of LRC instead of the flat dynamics of ARC. 
 
 However, imposing the constraints (\ref{match1}) sharply, i.e., setting $G^{\sigma,\sigma'}_\tau\! (j)=1$ if the constraints are satisfied, and $G^{\sigma,\sigma'}_\tau \!(j)=0$ otherwise, leads to a severe problem: as we allow only discrete values for the areas, the constraints (\ref{match1}) constitute diophantine conditions.  These conditions can only be satisfied  for a very small set of labels with accidental symmetries, e.g., if all 10 pairs of labels match \cite{Supp}. The resulting reduction in the density of states prevents a reasonable quantum dynamics.  This obstacle has also been encountered in higher gauge formulations of gravity \cite{BFCG1,Baratin,MV,BFCG2}.

One way out is to weaken the constraints (\ref{match1}), e.g., by allowing a certain error interval. But, one has to navigate between  Scylla---reducing too much the density of states---and Charybdis---imposing a dynamics that does not match GR.

 Here we will take guidance from  LQG and impose the constraints as strongly as allowed by the uncertainty relations resulting from (\ref{PB1}). To this end we  employ states  that are coherent in the two angle variables per tetrahedron, but restrict to the eigenspaces for the area operators. There are different constructions available for such states  \cite{Coherent, ConradyFreidel, BonzomLivine,FreidelHnybida}.
For a given tetrahedron $\tau$ we will denote  the coherent states ${\cal K}_\tau(\phi^{\tau}_{e_i};  \Phi_{e_i}^{\tau,\sigma})$, where $\phi^{\tau}_{e_i}=(\phi^{\tau}_{e_1},\phi^{\tau}_{e_2})$ are the arguments of the wave functions 
and $ \Phi_{e_i}^{\tau,\sigma}=(\Phi^{\tau,\sigma}_{e_1},\Phi^{\tau,\sigma}_{e_2})$  are the angles on which the wave function is peaked.  The coherent states come with a measure $d\mu^\tau_{\cal{K}}(\phi_1,\phi_2)$,  the precise form of which is immaterial here. These coherent states can be used to define the path integral 
 \ba\label{NewPI1}
 {\cal Z}'&=&\sum_{\{j_t\}} \mu(j) \!\!  \int \prod_\tau d\mu^\tau_{\cal{K}}(\phi)  \prod_{t } {\cal A}_t(j) \prod_\sigma {\cal A}'_\sigma(j,\phi)\, ,\;\;\;\;
 \ea
 where the new simplex amplitude is given by
 \ba
{\cal A}'_\sigma(j,\phi)= {\cal A}_\sigma(j) \prod_{\tau \in \sigma} {\cal K}_\tau(\phi^\tau_{e_i}; \Phi_{e_i}^{\tau,\sigma}(j)) \, .
 \ea
 Integrating out the angles we regain---without approximation and modulo boundary contributions \cite{boundCont}---a path integral of the form (\ref{Z1}) where the factors $G^{\sigma,\sigma'}_\tau$ are given by inner products between coherent states peaked on  the angles in $\tau$ induced by the areas of $\sigma$ and $\sigma'$, respectively,
 \ba\label{Gdef}
G^{\sigma,\sigma'}_\tau (j)= \langle  {\cal K}_\tau(\cdot ; \Phi_{e_i}^{\tau,\sigma}(j))\,|\, {\cal K}_\tau(\cdot ; \Phi_{e_i}^{\tau,\sigma'}(j)) \rangle \, .
 \ea
 By construction, this inner product is  peaked on the matching conditions (\ref{match1})  and provides a precise sense in which they are weakly imposed.  Imposition of these constraints leads to the brackets (\ref{PB1}), and through them to our main results (\ref{sc1}) and (\ref{sc2}). Counting the spin configurations contained in different confidence intervals of the $G^{\sigma,\sigma'}_\tau$ factors suggests that the weak imposition of the area constraints leads to a reasonable number of configurations contributing to the path integral (\ref{Z1}), see \cite{Supp}.

\medskip

\emph{On the flatness problem}.  We consider a first test case for the dynamics encoded in the path integral (\ref{Z1}). We choose a triangulation where we can control the scale for the bulk area variable and the bulk curvature through the boundary data. The complex consists of three 4-simplices sharing a single bulk triangle. There are no bulk edges, thus no bulk variables to sum over in LRC, and the bulk deficit angle is determined by the boundary lengths. Nonetheless in ARC, there is one bulk variable to sum over, which imposes a vanishing deficit angle for the internal triangle in the unconstrained theory. 

The amplitude of the path integral consists of two pieces: (i) an oscillatory phase factor, given by the exponentiated ARC action and (ii) the $G$-factors, which are peaked on the area constraints  (\ref{match1}) and decay exponentially. We employ scaling arguments and approximate these factors by gaussians with deviation,
\ba\label{wm1}
\sigma(\Phi) \sim  \sqrt{{\sin \alpha^{t,\tau}_v/(j_t+\tfrac{1}{2}) }} \, \, ,
\ea
 determined by the Poisson brackets (\ref{PB1}).

The semiclassical limit, as usually understood for spin foams,  amounts to a large $j$ limit. The reason for this is the linear scaling of the action (\ref{ARegge}) in the areas, and thus in the spins $j$.
 As the amplitudes include an oscillatory  and an exponentially decaying factor, there are two conditions for configurations to contribute \cite{SFLimit,FlatProblem}: (i) the oscillatory phase should be stationary and (ii) the $G$-factors should be near their maxima.  As the oscillatory phase is given by the exponentiated Area Regge action, the stationarity condition (i) leads to flatness in the unconstrained model.  A similar structure of the amplitudes, and hence a similar flatness problem \cite{FlatProblem,Bonzom1,HellK, Han1,Oliveira1, DGS, ILQGS}, arises for the EPRL-FK spin foam models \cite{NSFM}.

In the following we will perform a more detailed analysis and identify a regime in which curved configurations can dominate. A bound, that has heretofore received little attention, emerges. We distinguish the spin scales set by the boundary, $j$, and by the bulk, $j_{\text{blk}}$.  For the minimal triangulations investigated here $j \sim j_{\text{blk}}$.

Since we are interested in the semiclassical limit, we focus on the classical action and on the exponent of the gaussian $G$-factors.  The small $\hbar$ limit of these exponents are dominated by the classical values, from which our scaling results will follow. From (\ref{wm1}) we see that the $G$-factors come with a deviation $\sigma(\Phi)\sim 1/ \sqrt{j}$ for the 3D dihedral angle, where we assume that the boundary areas $ a \sim \gamma\ell_P^2 j$ have approximately equal values, so that the bulk area scales as $a_{\text{blk}}\sim a$. As angles are dimensionless, their derivatives  scale as $\partial \Phi/\partial a_{\text{blk}}\sim 1/a$ and $\partial \epsilon /\partial a_{\text{blk}}\sim 1/a$. Thus, we have $\partial \Phi/\partial j_{\text{blk}} \sim 1/j$ and $\partial \epsilon /\partial j_{\text{blk}} \sim 1/j$, and the deviations  of the gaussian $G$-factors, expressed as functions of bulk spin and deficit angle $\epsilon$, respectively, scale as 
\ba\label{sc1}
\sigma(j_{\text{blk}})&\sim&  \left[\frac{\partial \Phi(j_{\text{blk}} )}{\partial j_{\text{blk}} } \right]^{-1}\!\!\!\times\sigma(\Phi) \,\, \sim \, \, j \times \frac{1}{\sqrt{j}} \,=\, \sqrt{j} \, ,\nn\\
\sigma(\epsilon)&\sim& \left[\frac{\partial \epsilon(j_{\text{blk}} )}{\partial j_{\text{blk}} } \right]\times\sigma(j_{\text{blk}}) \,\,\sim\, \,\frac{1}{j} \times \sqrt{j}=\frac{1}{\sqrt{j}} \, .\,\,\,\,\;\,
\ea
As the angles are invariant under rescaling, we can choose boundary data that induce a given deficit angle $\epsilon$, and then choose a sufficiently large scale $j$, so that the $\epsilon=0$ value is outside the deviation interval. Thus by going to sufficiently large spins $j$, the constraint part of the amplitudes can peak sharply on non-vanishing curvatures. Note that the  deviations $\sigma(j_{\text{blk}})$ and $\sigma(\epsilon)$  as functions of $j$ are independent of the spectral-spacing parameter $\gamma$.

\begin{figure}[!t]
   \centering
    \begin{subfigure}[b]{0.2291\textwidth}
        \includegraphics[width=\textwidth]{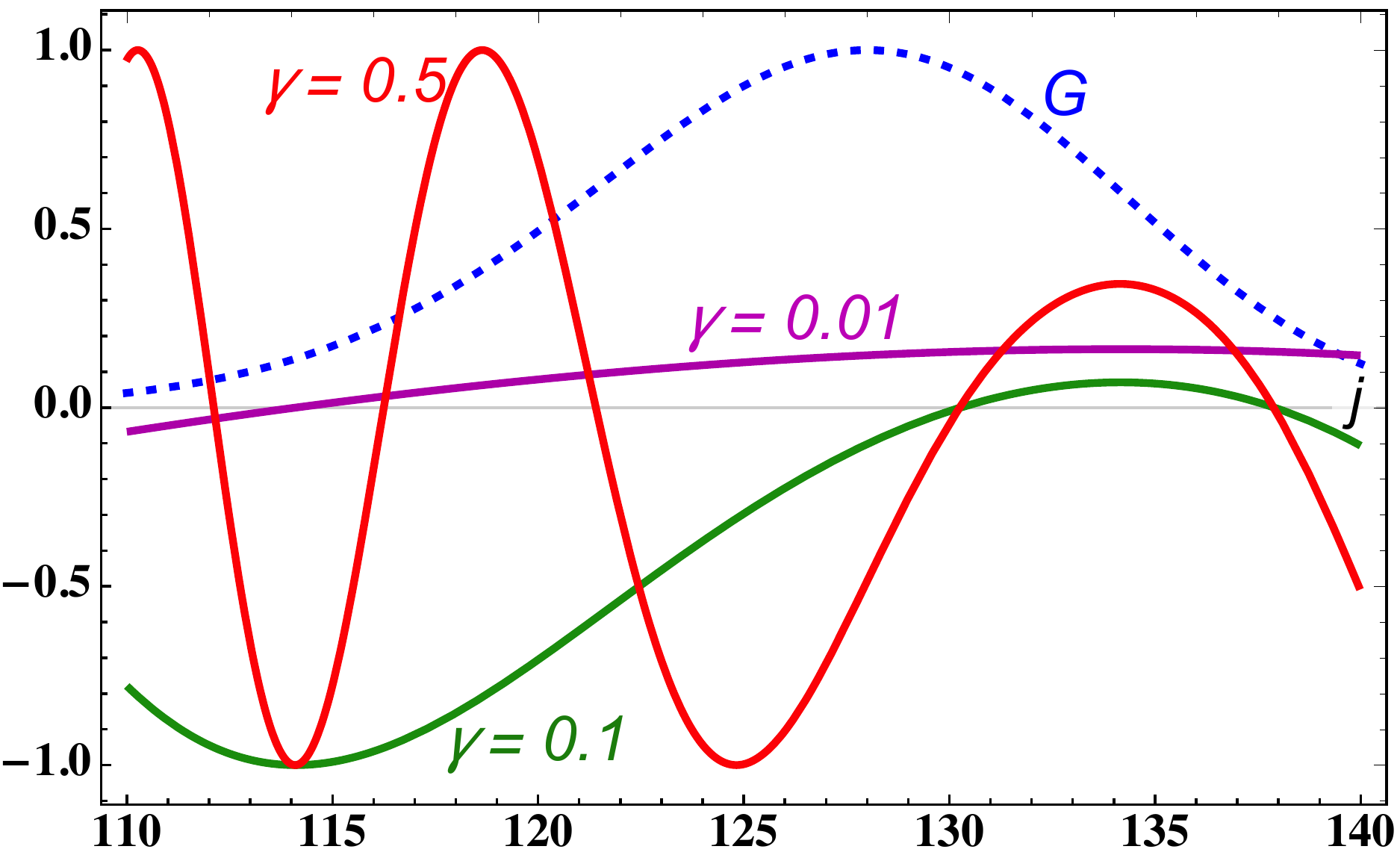}
        \caption{$j=99.5$ }
    \end{subfigure}
    \quad
    \begin{subfigure}[b]{0.2291\textwidth}
        \includegraphics[width=\textwidth]{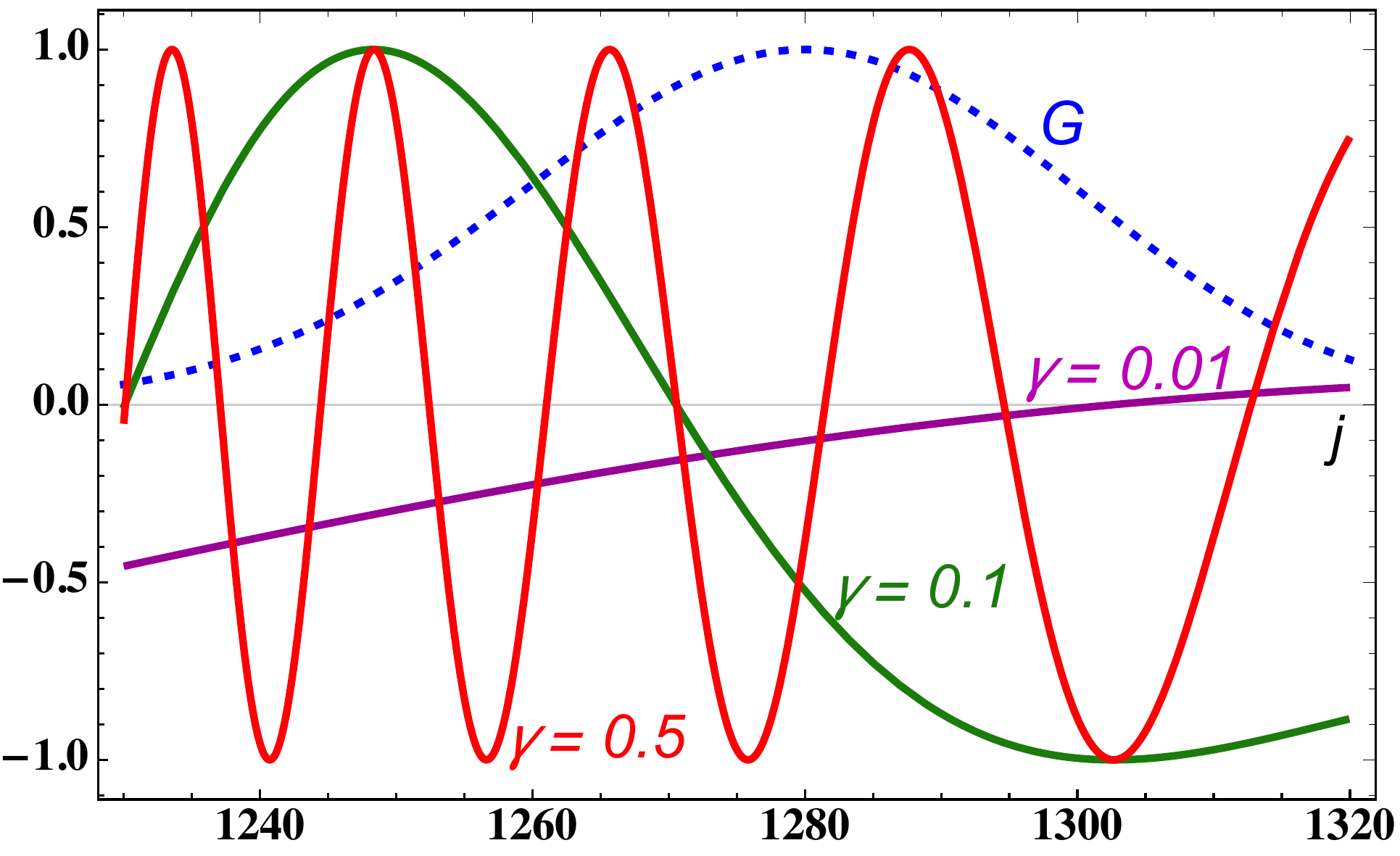}
        \caption{$j=999.5$}
    \end{subfigure}    
    \caption{The $G$-factors (dashed), which impose the matching conditions weakly,  and the real part of the product of the amplitude factors ${\cal A}_t$ and ${\cal A}_\sigma$  as functions of the bulk spin $j_{\rm blk}$ for various $\gamma$ (solid lines). Here, the $G$ factors peak on a curvature value $\epsilon\approx 0.5$. Larger $\gamma$'s lead to a more oscillatory behaviour.  These examples are detailed in the supplemental material \cite{Supp}.}\label{LambdaPhiM}
\end{figure}

Although the $G$-factors can be peaked on curved configurations, the relevant summation range for the bulk spin $j_{\rm blk}$ scales with $\sqrt{j}$. The oscillatory phase factor can, therefore, average out the expectation value for the deficit angle,  see Fig. \ref{LambdaPhiM}. To avoid this, we need to make sure the oscillations are sufficiently slow
\ba\label{sc2}
\sigma( \tfrac{S_{\rm ARC} }{\ell_P^2} )=\frac{1}{\ell_P^2} \frac{ \partial (S_{\rm ARC})}{\partial j_{\text{blk}}} \!\times \!\sigma( j_{\text{blk}} ) \sim \gamma \epsilon \sqrt{j} \stackrel{!}{\lesssim} {\cal O}(1)\,.\;\;\;\;
\ea
Thus, whereas the scaling for the deficit angle (\ref{sc1}) requires a choice of larger $j$, (\ref{sc2}) demands that with growing $j$ we choose smaller $\gamma$, and thus a smaller spacing between the area eigenvalues. Taking $j$ large and keeping $\gamma$ fixed---as for the large $j$-limit discussed above  and often treated in the literature---the phase factor will oscillate more and more strongly and suppress the configurations on which the $G$-factors are peaked, see  Fig. \ref{LambdaPhiM}. These expectations are confirmed by numerical examples in the supplemental material \cite{Supp}. 

We note that $\gamma$ enters  (\ref{sc2}) simply because $S_{\rm ARC}$ is linear in the $a$ and $\partial a/\partial j \sim \gamma$ from \eqref{ASpec}.   Depending on which quantity we consider fixed,  we can also interpret  (\ref{sc2}) as a bound on  the curvature per triangle $\epsilon \lesssim 1 /(\gamma \sqrt{j})$, or  a $\gamma$- and curvature dependent upper bound on the spin  $j$.

We have considered the simplest triangulation that differentiates between LRC and ARC. 
 As we only employ scaling arguments, similar conclusions may also apply for larger triangulations.  In future work we will investigate examples including bulk edges and vertices.  Finally, to reach definite conclusions on the continuum limit it will be necessary to see how the implementation of the constraints changes under coarse graining.  The models proposed here simplify this task immensely. 

\medskip

\emph{Discussion}.
Area operators are central in a number of approaches to 4D  quantum gravity,  notably LQG and holography. To achieve a quantum dynamics that reproduces GR, constraints between the areas need to hold. This is, however, hindered if areas have an asymptotically equispaced spectrum and are locally independent.  

The imposition of these constraints is pivotal in spin foam quantization. This leads  to involved amplitudes, which has so far prevented satisfactory resolution of key dynamical questions, most pressingly whether the models suppress curvature excitations. Thus we propose to use, instead of the standard spin foam amplitudes, a class of effective models, with a  transparent encoding of the dynamics and much more amenable for numerical investigations \cite{Dona:2019dkf,Dona:2020}.  If it becomes clear that these effective models lead to a gravitational dynamics in the large scale limit, one can study the effects of using more involved versions, including a sum over orientations or degenerate geometries \cite{Engle}. See \cite{3DHol} for first insights in the 3D case.

 In the models proposed here the constraints are imposed  weakly, but as strongly as allowed by the LQG Hilbert space,  from which the discrete,  locally independent 
 area spectra result.  Whether this  leads to  the correct dynamics is not understood, even in much simpler models than gravity, and should be further tested. In particular, for spin foam models, a too weak imposition of the constraints could lead to suppression of curvature.

We have found that curvature is not necessarily suppressed. This result comes with restrictions connecting the average area $a \sim  \ell_P^2 \gamma j$, the Barbero-Immirzi parameter $\gamma$, and the curvature $\epsilon_t$ per triangle. The concentration of the constraints  on a given curvature value improves with growing spin $j$, as $1/\sqrt{j}$, but is independent of $\gamma$.  Our condition $\gamma \sqrt{j} \epsilon_t \lesssim {\cal O}(1)$, prefers small  $\gamma$, and hence a small spacing in the area spectrum. 

In numerical examples \cite{Supp}, we need large spins and small $\gamma$ to obtain an expectation value for the deficit angle that approximates well the classical value. This justifies our focus on `effective' models, where we replace the full spin foam simplex amplitude with its large spin asymptotics, which is already obtained in practice around $j=10$, and is given by the cosine (replaced here with the exponential) of the Regge action \cite{BC, SFLimit,SFtimelike,SFPoly}. 

 It has been argued in \cite{CE}, that a double scaling limit that takes $\gamma$ small and spins $j$ large, with $\gamma j$ fixed, reproduces the LRC equations of motion. Here, we find also that $\gamma$ should be small and $j$ large, but that we need for the combination $\gamma \sqrt{j} \epsilon_t$ to be of order one or smaller. This combination, and the related bound on curvature has also been identified in \cite{MHan}, based on a generalized stationary phase analysis of the EPRL-FK amplitudes. Using much simpler inputs, we have shown that this bound  does not depend on specific choices for the spin foam amplitudes.  Rather, the reason for this bound is tied to the LQG Hilbert space and the area spectrum it leads to. 

The conclusions for the expectation value of the curvature hold in general, but assume that we 
can  control the scale of bulk spin and deficit angles, e.g., via the choice of boundary data. This is not necessarily the case for larger triangulations. To understand the continuum limit, we will have to investigate how these arguments are impacted by coarse graining and renormalization \cite{CG}. 
The investigation of corresponding continuum actions \cite{Krasnov}, in which the geometricity constraints  are also imposed weakly, might elucidate how these constraints behave under renormalization.

The effective model presented here is the numerically fastest spin foam in the literature to date. All the computations for this paper were performed on individual laptops. The recent work \cite{Dona:2020}, which uses the same triangulation, but works with 4D BF-theory was computed on 32-core machines. No comparable computation has been carried out for the full EPRL-FK models \cite{NSFM}.  Effective spin foams should make the study of coarse graining flow \cite{CG} more feasible than for other spin foam models and will help to establish whether LQG and spin foams allow for a satisfactory continuum limit.

\smallskip

\begin{acknowledgments}
\emph{Acknowledgments}. BD thanks Wojciech Kaminski, and BD and HMH thank Abhay Ashtekar, Eugenio Bianchi, Pietro Don\`a, Aldo Riello, and Simone Speziale for discussions. SKA is supported by an NSERC grant awarded to BD. HMH gratefully acknowledges support from the visiting fellows program at the Perimeter Institute and the warm hospitality of the quantum gravity group. 
Research at Perimeter Institute is supported in part by the Government of Canada through the Department of Innovation, Science and Economic Development Canada and by the Province of Ontario through the Ministry of Colleges and Universities.
\end{acknowledgments}

\section{-- Supplemental Material -- \\Effective Spin Foam Models for Four-Dimensional Quantum Gravity}

\renewcommand{\thefigure}{A\arabic{figure}}

\setcounter{figure}{0}

 This supplement consists of a proof of Eq. \eqref{PB1} from the main text (simpler than the one that is currently present in the literature), numerical investigation of the length configurations as the number of discrete area parameters is increased, and  a numerical investigation of the simplest triangulation containing a bulk triangle. Equations in the main text are referred to with standard numerals, while those in the supplemental material are preceded by an $S$. 

\medskip

\emph{Derivation of the brackets between dihedral angles}.  The Poisson bracket on $N$ copies of the dual to the Lie algebra of $\mathfrak{su}(2)$ is 
\begin{equation} \tag{S1}
\label{brack}
\{ f, g \} = \sum_{I =1}^{N} \vec{J}_{I} \cdot \left( \frac{ \partial f}{\partial \vec{J}_I} \times \frac{\partial g}{\partial \vec{J}_{I}} \right), 
\end{equation}
where each of the $\{\vec{J}_{I}\}_{I=1}^{N}$ can be thought of as an angular momentum vector $\vec{J_I}=(J_{I1},J_{I2},J_{I3})$. 
Consider the case $N=4$, which gives the phase space of a tetrahedron described in terms of the Minkowski area vectors $\vec{A}_I=\gamma \vec{J}_I$ whose magnitudes $A_I$ are equal to the tetrahedron's face areas and whose directions are normal to the faces. (References \cite{BianchiHaggard1} provide additional details about this phase space and how it connects to the geometry of the tetrahedron.) The  (internal) dihedral angle $\phi_{IK}$ between the triangles labelled  by $I$ and $K$ is determined by 
\ba
\vec{A}_{I} \cdot \vec{A}_K = - A_I A_K \cos \phi_{IK}. \nn
\ea

Define $B_{12,23} \equiv \{ \vec{A}_1 \cdot \vec{A}_2 , \vec{A}_2 \cdot \vec{A}_3 \}$ and note that 
\be \tag{S2}
\label{angExp}
B_{12,23}
= A_1 A_2^2 A_3 \sin \phi_{12} \sin \phi_{12} \{  \phi_{12},  \phi_{23} \} , \quad \, \, 
\ee
where we used the derivation property of the bracket. On the other hand, we can make use of Eq. \eqref{brack} to compute
\begin{align}
\label{Bcalc}
B_{12,23}&= \sum_{I} \epsilon_{ijk} J_{I i } \frac{\partial}{\partial J_{I j} } (A_{1m} A_{2m}) \frac{\partial}{\partial J_{I k} } (A_{2n} A_{3n}) \nn \\
& = - \gamma\, \, \vec{A}_{1} \cdot (\vec{A}_2 \times \vec{A}_3) \,=\,
 -\gamma\, \frac{9}{2}V^2  \tag{S3}
\end{align}
where we drew on the following expression for the volume of a tetrahedron $V^2 = \tfrac{2}{9} \vec{A}_{1} \cdot (\vec{A}_2 \times \vec{A}_3).$
This volume can be also computed through $ V = \tfrac{2}{3} l^{-1}_{IK}  A_I A_K \sin \phi_{IK},$ where $l_{IK}$ is the length of the edge shared by triangles $I$ and $K$. Using this formula twice gives
\begin{equation} \tag{S4}
\label{volExp}
B_{12,23} = - \frac{2\gamma}{l_{12} l_{23} } A_1 A_2^2 A_3 \sin \phi_{12} \sin \phi_{23}.
\end{equation}
Setting the two expressions \eqref{angExp} and \eqref{volExp} equal gives
\[
\{  \phi_{12},  \phi_{23} \}  = - \frac{2\gamma}{l_{12} l_{23}} .
\]

Finally, noting that 
$
A_2 = \tfrac{1}{2} l_{12}l_{23} \sin \alpha_{12, 23},
$
where $\alpha_{12, 23}$ is the angle between the edges with lengths $l_{12}$ and $l_{23}$, yields the result quoted in the text, Eq. \eqref{PB1},
\begin{equation} \tag{S5}
\{  \phi_{12},  \phi_{23} \}  = -\gamma \frac{\sin \alpha_{12, 23}}{A_2} .
\end{equation}
Of course, the minus sign in this result is just due to the choice of an increasing index ordering for the bracket. 

\medskip

\begin{figure}[!h]
   \centering
    \begin{subfigure}[b]{0.2294\textwidth}
        \includegraphics[width=\textwidth]{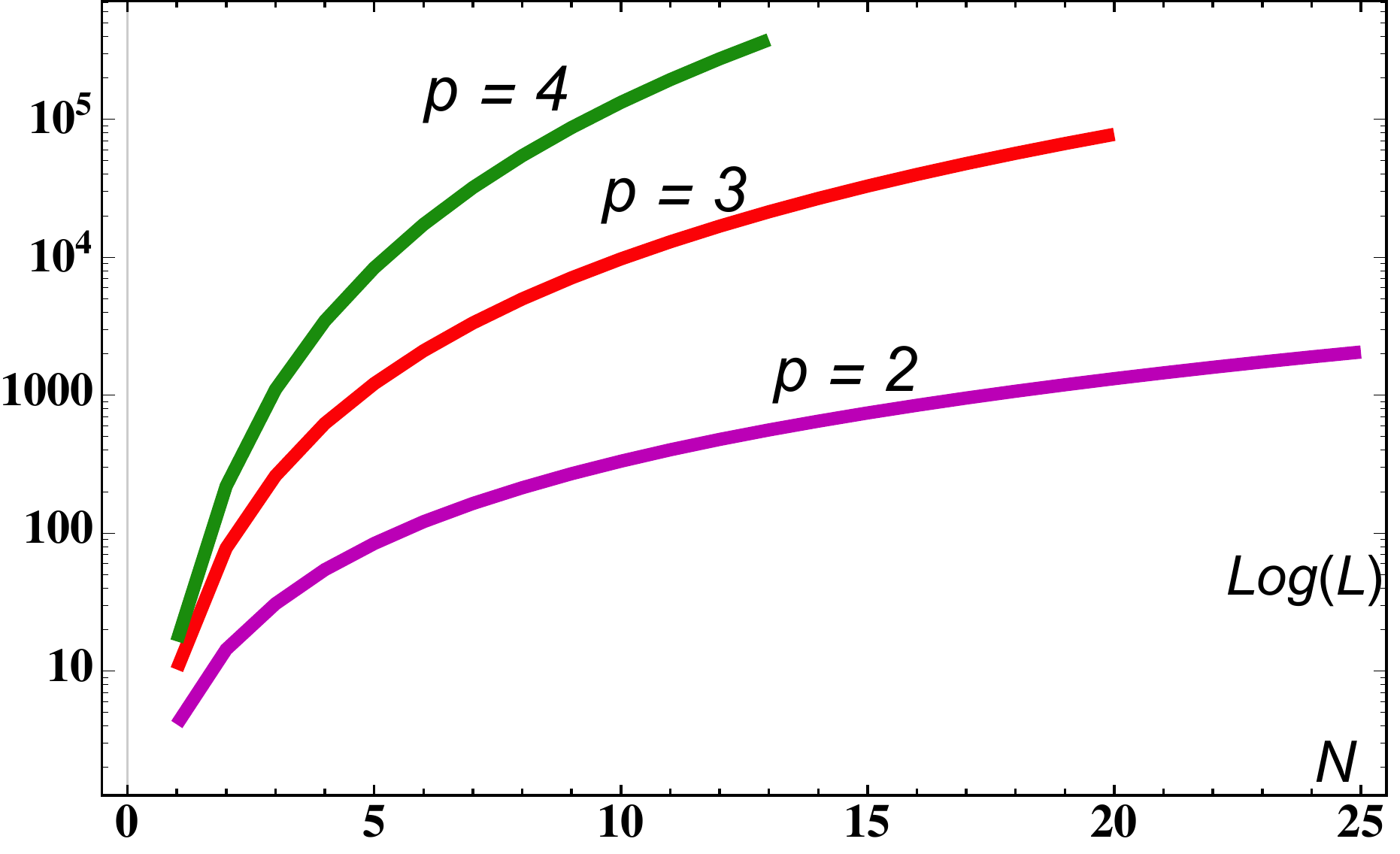}
        \caption{}
        \label{lconf1}
    \end{subfigure}
    \, 
    \begin{subfigure}[b]{0.2294\textwidth}
        \includegraphics[width=\textwidth]{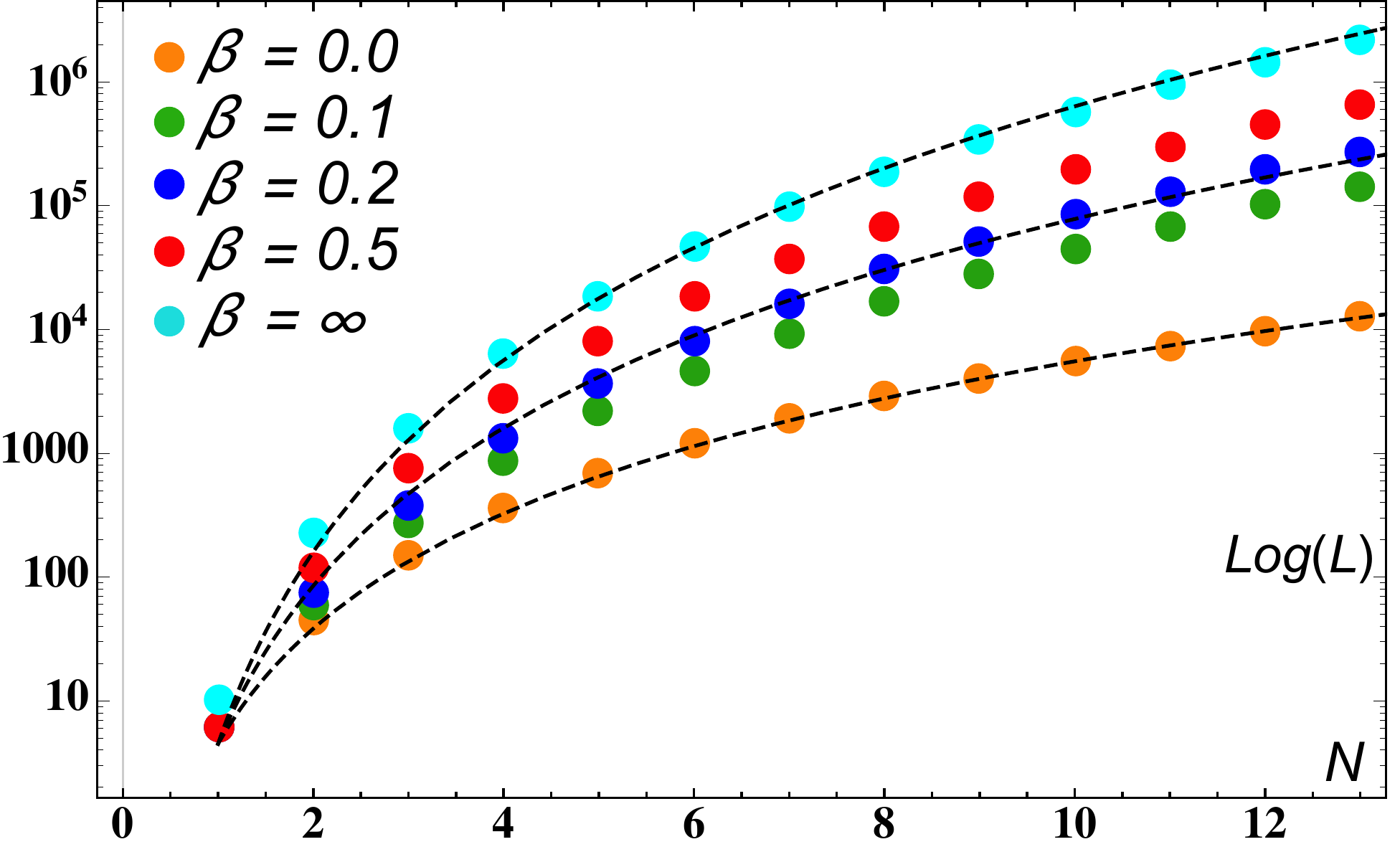}
        \caption{}
        \label{lconf2}
    \end{subfigure} 
    \vspace{-0.25cm}
    \caption{Log-linear plots of the number $L$ of length configurations  as a function of the maximal area. (a) Count of length configurations with areas up to $N$ in a simplex with $p$ length parameters. (b) The count for two glued simplices with four length parameters. The dashed lines show $N^3$, $N^4$ and $N^5$ power law scaling. }\label{LambdaPhi0}
\end{figure}

\emph{Counting of length configurations}. We consider a triangulation with certain edge lengths chosen to be equal and then compute the number of allowed edge length solutions given locally independent discrete asymptotically equidistant area spectra. In order to see how the number of allowed configurations scales if there are no area constraints we consider just one 4-simplex with vertices (12345) and $p \in \{2,3,4\}$ length parameters. For $p=2$, we set $l_{ij}=x$ and $l_{i5} = y$, with $i,j \in \{1,2,3,4\}$. For the $p=3$ case we choose: $l_{ij} = x, l_{mn}=y $ and $l_{im} = z$ with $i,j\in \{1,2,3\}$ and $m,n\in \{4,5\}$. For $p=4$ we have $l_{ij} = w, l_{i4}=x,l_{i5}=y $ and $l_{45}=z$ with $i,j \in \{1,2,3\}$. We count all edge length solutions where the triangle areas take discrete values $A_t \in \{ \tfrac12, 1, \cdots ,N\}$ for $N \in \mathbb N$. The left panel of Figure \ref{LambdaPhi0} shows a semilog plot of the number of length solutions for a simplex having $p \in \{2,3,4\}$ length parameters. The number of length configurations scales as $ N^{1.03p} \approx N^p$. \\

\vspace{-0.25cm}

We also consider a gluing of two  simplices with vertices $\sigma=(12345)$ and $\sigma'=(12346)$. For the shared tetrahedron we allow two length parameters $u$ for the edges $(12)$ and $(34)$ and $v$ for the remaining four edges. All four areas of the tetrahedron therefore agree, and we are left with one area parameter $a=A(u,v,v)$. Here $A(x_1,x_2,x_3)$ denotes the area of a triangle with edge lengths $(x_1,x_2,x_3)$. For the simplex $\sigma$ we introduce additional edge lengths $w$ for edges $(i5)$ with $i \in \{1,2,3,4 \}$. This introduces two more area parameters $b=A(u,w,w)$ and $c=A(v,w,w)$, giving us three length and three area parameters for $\sigma$. We make the same kind of choices for $\sigma'$, that is, $w'$ gives the length of the edges $(i6)$ leading to area parameters $b'=A(u,w',w')$ and $c'=A(v,w',w')$.  After gluing the complex has four length parameters $(u,v,w,w')$ and five area parameters $(a,b,c,b',c')$. 

We proceed to count the number of configurations with all areas valued in  $\{ \tfrac12, 1, \cdots ,N\}$, and with maximum deviations given by
\be \tag{S6}
\sigma(\Phi) =\beta \sqrt{\ell^2_P \gamma \frac{\sin \alpha^{t,\tau}_v}{a_t}} = \beta\sqrt{\frac{\sin \alpha^{t,\tau}_v}{(j_t+\tfrac{1}{2}) }} 
\ee
for the pairs $(\Phi^{\tau,\sigma}_{e_i},\Phi^{\tau,\sigma'}_{e_i})$  of 3d dihedral angles in the shared tetrahedron $\tau$.  This is informed by the deviation Eq. \eqref{wm1} for the $G$-factors.  Explicitly we implement the deviation constraints using a Heaviside approximation to the $G$-factors
\ba
|\Phi^{\tau,\sigma}_{e_i}-\Phi^{\tau,\sigma'}_{e_i}|<\sigma(\Phi), \quad i\in\{1,2\}. \nn
\ea

 We have also introduced a dimensionless parameter $\beta$ to be able to tune between an exact imposition ($\beta=0$) and no imposition ($\beta=\infty$) of the matching constraints.
The right panel of Fig. \ref{LambdaPhi0} shows the results for various choices of $\beta$.  For $\beta=0$ we find a scaling $N^{3}$. This is explained by the fact that requiring exact matching forces $w=w'$, and thus we have only three parameters. Not imposing the matching conditions, we find a scaling $N^{5}$ reflecting the five area parameters for the two glued simplices. For $\beta \approx 0.15$ we find a scaling of $N^{ 4}$, see Fig.~\ref{LambdaPhi0},   which is the same as we found for one simplex with $p=4$ free length parameters. 

\medskip

\emph{Triangulations with three and with six 4-simplices}. Take three 4-simplices with  vertices $(12345),(12356)$ and $(13456)$ respectively, and  glue these around the  shared triangle $(135)$. Here all edges and all but the triangle $(135)$ are in the boundary. Thus we have one bulk triangle and no bulk edges. 

We will assume some lengths to be equal, so that we have overall only three length parameters: $x=l_{ij}$, $y=l_{mn}$, and $z=l_{im}$, with $i,j \in \{1,3,5\}$ and $m,n \in \{2,4,6\}$. Correspondingly, we have three area parameters $a=A(x,x,x)$, $b=A(x,z,z)$, and $c=A(y,z,z)$ where $A(x_1,x_2,x_3)$ denotes the area of a triangle with lengths $(x_1,x_2,x_3)$.

Note that with this special choice of boundary data the boundary areas $(b,c)$ do not determine the boundary lengths $(x,y,z)$. To do so one also needs the bulk area $a$. In Area Angle Regge calculus one has also 3D dihedral angles as boundary data. With the given symmetry reduction, all boundary tetrahedra have the same geometry, determined by edge lengths $(z,y,z,z,x,z)$. We can choose a pair of non-opposite edges, both with length $z$. Due to our  choice of symmetric boundary data, the 3D dihedral angles $\phi_z$ for the $z$-edges are all the same---thus we have boundary data $(b,c,\phi_z)$. These determine a bulk deficit angle $\epsilon_a(b,c,\phi_z)$.

The matching conditions Eq. \eqref{match2} for the bulk tetrahedra are all satisfied due to our symmetry reduction. Thus, if we start from the  path integral with area and angle variables Eq. \eqref{NewPI1}, and integrate out the bulk 3D dihedral angles, we will just obtain a multiplicative factor, given by the norm of the coherent states ${\cal K}_\tau(\cdot, \Phi)$.

We can now consider this path integral with a boundary, which, after integrating out the bulk 3D  angles, involves only a summation over one  spin $j_a$. 

Alternatively, we can take two such complexes consisting of three 4-simplices each, and glue these so that we obtain a triangulation of $S^4$. After integrating out all 3D dihedral angles we will have four area parameters, the areas $b$ and $c$ and the bulk areas $a$ and $a'$ from the two complexes respectively. We will  compute the  expectation value for the deficit angle $\langle\epsilon_a\rangle$---while keeping the areas $(a',b,c)$ fixed. Classically, i.e., with sharp shape-matching constraints and fixed $(a',b,c)$, these data  determine the deficit angles $\epsilon_{a'}$ and $\epsilon_a$ with $\epsilon_a=\epsilon_{a'}$.

The summation for the path integral thus involves only the bulk area parameter $a$. There are two contributions to the amplitudes: the exponential of the (Area) Regge action, as well as the inner product $G(a,a')$ between the coherent states, which impose the matching constraints $\Phi_z(a,b,c)=\Phi_z(a',b,c)$. 
We approximate the factor arising from these inner products between the coherent states by a gaussian
\be\tag{S7}
\label{S7}
G(a,a')=\exp\big( -\frac{9}{2  \sigma^2(\Phi) } (\Phi_z(a,b,c)-\Phi_z(a',b,c))^2\big)\nn
\ee
with a deviation $\sigma$ determined by the Poisson brackets \eqref{PB1}:
\be \tag{S8}
\label{S8}
 \sigma^2(\Phi)=\frac{1}{2}\frac{\sin \alpha(a,b,c)}{(j_b+1/2)}+\frac{1}{2}\frac{\sin \alpha(a',b,c)}{(j_{b}+1/2)}\, ,
\ee
where $\sin \alpha(a,b,c)=2 \sqrt{3} a b /(a^2+3b^2)$. The factor 9 in the exponential arises because we have 9 boundary tetrahedra and therefore 9 inner products.

For the computation of the expectation value $\langle \epsilon_a\rangle(a',b,c)$ we use
\be \tag{S9}
\langle \epsilon_a\rangle(a',b,c)= \frac{1}{{\cal Z}}\sum_{j_a}  \epsilon_a \,\, G(a,a') \prod_t {\cal A}_t \prod_\sigma {\cal A}_\sigma 
\ee
with 
\be \tag{S10}
{\cal Z}=\sum_{j_a}  G(a,a') \prod_t {\cal A}_t \prod_\sigma {\cal A}_\sigma
\ee
and ${\cal A}_t$ and ${\cal A}_\sigma$ defined below Eq. \eqref{Z1} of the main text. 

We use homogeneous boundary areas, that is, choose $b=c$. In this case the angles $\Phi_z$ and $\epsilon_a$ are given by 
\be\tag{S11}
\Phi_z=\arccos\left( \frac{a^2}{3b^2}\right) , \qquad \text{and} \qquad  
\ee
\be\tag{S12}
\epsilon_a=2\pi-3 \arccos\left(\frac{9b^2-7a^2}{9b^2-a^2}\right)\, ;
\ee
note that triangle inequalities imply $a^2\leq 3b^2$. The area action for the complex with three simplices is
\ba
S&=&9b\left[\frac{a}{9b} \epsilon_a+2\pi  - \arccos \left(\frac{3b^2-2a^2}{6b^2-2a^2} \right)  \right. \nn\\
&&\left. \hspace{2.2cm} -\arccos \left(\frac{a^2}{\sqrt{27b^4-12a^2b^2+a^4}}\right)\right] \, ,\nn
\ea
and the $G$-function is given in \eqref{S7} and \eqref{S8} with $b=c$.

The resulting expectation values are shown in Tables \ref{TableEps1} and \ref{TableEps2}. Here we have set $j_b=j_c=j$. Thus the pair $(j,j_{a'})$ determine the scale as well as the deficit angle $\epsilon_{a'}$. Classically we have $\epsilon_{a}=\epsilon_{a'}$. To reproduce this result for the expectation value we need a sufficiently large scale $j$ and a sufficiently small value for the Barbero-Immirzi parameter $\gamma$, in particular if we consider data leading to a small deficit angle.

 \begin{table}[h!]
  \begin{center}
  $
\begin{array}{|c|c|c|c|}\hline
(j+\tfrac{1}{2},j_{a'}+\tfrac{1}{2},\epsilon_{a'}) &\gamma=0.01&\gamma=0.1 &\gamma=0.5\\\hline
(30,38.5,0.52)&0.78-0.03{\rm i}&0.68-0.26{\rm i}&0.17  -0.32{\rm i} \\
(100,128,0.54)&0.62-0.062{\rm i}&0.55-0.19{\rm i}&0.17  -0.27{\rm i} \\
(300,384,0.54)&0.57-0.02{\rm i}&0.51-0.17{\rm i}&0.16  -0.25{\rm i} \\
(1000,1280,0.54)&0.55-0.01{\rm i}&0.50-0.16{\rm i}&0.16  -0.24{\rm i}     \\\hline
\end{array}
$
  \caption{Expectation value for the deficit angle $\epsilon_a$ with classical value $\approx 0.5$ for various $j$, $j_a'$, and $\gamma$. }
  \label{TableEps1}
\end{center}
 \end{table}


 \begin{table}[h!]
  \begin{center}
  $
\begin{array}{|c|c|c|c|}\hline
(j+\tfrac{1}{2},j_{a'}+\tfrac{1}{2},\epsilon_{a'}) &\gamma=0.01&\gamma=0.1 &\gamma=0.5\\\hline
(30,40,0.08)&0.39-0.02{\rm i}&0.33-0.15{\rm i}&0.03  -0.14{\rm i} \\
(100,133.5,0.06)&0.14-0.01{\rm i}&0.13-0.05{\rm i}&0.03  -0.06{\rm i} \\
(300,400,0.08)&0.11-0.00{\rm i}&0.09-0.03{\rm i}&0.03  -0.5{\rm i} \\
(1000,1335,0.06)&0.07-0.00{\rm i}&0.06-0.02{\rm i}&0.02 -0.03{\rm i}     \\\hline
\end{array}
$
  \caption{Expectation value for the deficit angle $\epsilon_a$ with classical value $\approx 0.07$ for various $j$, $j_a'$, and $\gamma$. }
  \label{TableEps2}
\end{center}
 \end{table}

\begin{figure}[h]
   \centering
    \begin{subfigure}[b]{0.2291\textwidth}
        \includegraphics[width=\textwidth]{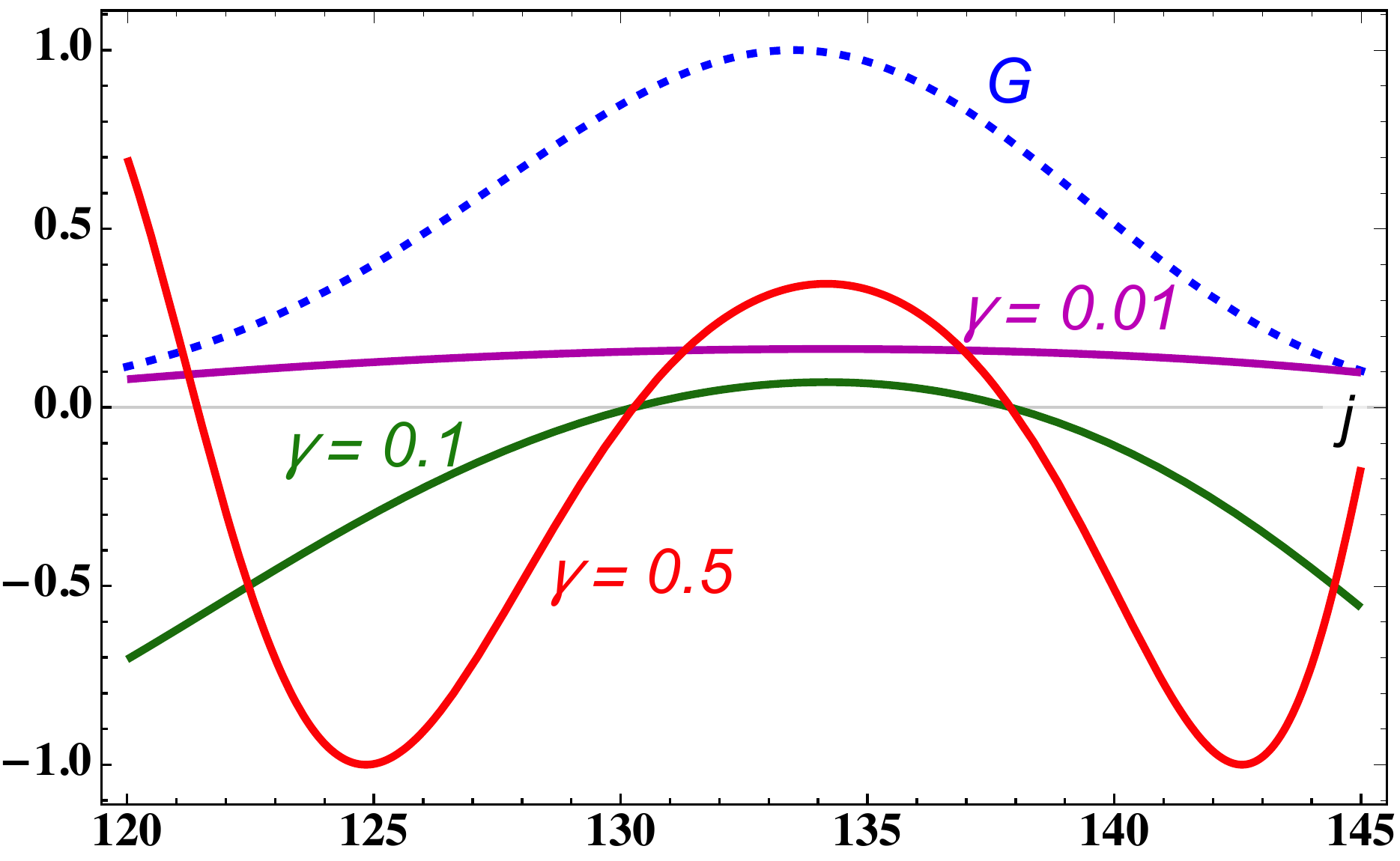}
        \caption{$j=99.5$ }
        \label{grph1}
    \end{subfigure}
    \quad
    \begin{subfigure}[b]{0.2291\textwidth}
        \includegraphics[width=\textwidth]{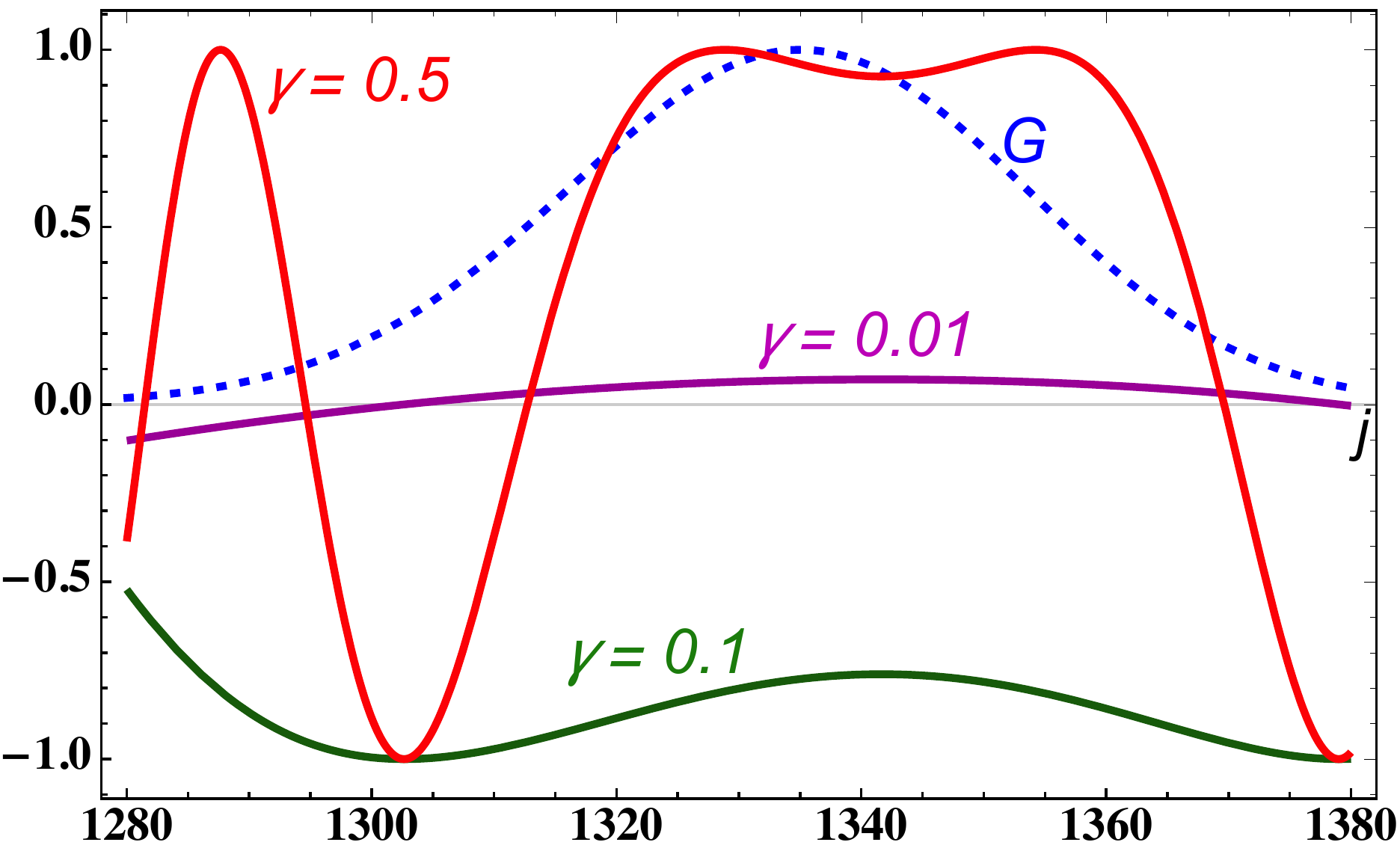}
        \caption{$j=999.5$}
        \label{prph2}
    \end{subfigure}    
    \caption{The $G(a,a')$ factor (dashed) and the real part of the product of the amplitude factors ${\cal A}_t$ and ${\cal A}_\sigma$ as a function of $j_a$ for $\epsilon_{a'}\sim0.07$ and  different $\gamma$--values.
    }\label{LambdaPhi}
\end{figure}

 In this example the averaging of the deficit angle with the $G(a,a')$ factor (but without the ${\cal A}_t$ and ${\cal A}_\sigma$ factors) tends to over-estimate the curvature angle. This is due to a certain asymmetry in the example that partially originates with the generalized triangle inequalities, which restrict $a$ to $a\leq \tfrac{3}{2}b=\tfrac{3}{2}c$.  The  oscillatory behavior of the ${\cal A}_t$ and ${\cal A}_\sigma$ factors tends to average out the expectation values---more so for larger Barbero--Immirzi parameter $\gamma$, which leads to more oscillations over the interval where $G(a,a')$ is sufficiently large, see Fig.~\ref{LambdaPhi} and Fig.~1 (in the main text). Note that the expectation values do have imaginary contributions.  These arise as the $G(a,a')$ factor peaks away from the stationary point of the action (where $\epsilon_a=0$), so the imaginary parts do not average out. As the imaginary contributions  are sourced by the oscillatory behaviour of the amplitudes, they grow with $\gamma$. Having imaginary contributions on the order of the real contributions indicates that the regime is unreliable, even if the (real part of the) expectation value happens to be near the classical value.

\providecommand{\href}[2]{#2}
\begingroup
\endgroup


\begin{thebibliography}{100}
\bibitem{inversion}
 These functions depend on a discrete parameter that accounts for multiple roots appearing in this inversion. This parameter is another summation variable for the (constrained) ARC path integral. To simplify notation we suppress it here.
\bibitem{noncommutative}
 This Poisson bracket (with $\gamma=1$) appears in the Kapovich-Millson phase space \cite{KM}, which is used to describe the space of shapes of tetrahedra with fixed areas \cite{ConradyFreidel,Twisted, BianchiDonaSpeziale,BianchiHaggard1}. The non-commutativity of the angles is inherited by the lengths  \cite{BianchiLength,DittrichRyan}.
\bibitem{boundCont}
  These are given by a coherent state ${\cal K}_\tau$  for each boundary tetrahedron.
\end{thebibliography}
\end{document}